\documentclass[aps, prl, twocolumn,twoside,superscriptaddress,longbibliography,floatfix,balancelastpage]{revtex4-2}

\usepackage[T1]{fontenc}

\usepackage{graphicx} % Required for inserting images
\usepackage{hyperref}
\usepackage{comment}
\usepackage{xcolor}
\usepackage{amsmath}
\usepackage{braket}

\definecolor{trackgreen}{rgb}{0.0,0.45,0.0}

\definecolor{trackmagenta}{rgb}{0.65,0.0,0.45}

\begin{document}

\title{The Critical Role of Itinerant Contributions to Orbital Angular Momentum Relaxation and Dynamics}

\author{Andrew C. Grieder}
\thanks{A.C.G.\ and L.M.C.\ contributed equally.}
\affiliation{Department of Materials Science and Engineering, University of Wisconsin-Madison, 53706, United States}

\author{Luis M. Canonico}
\thanks{A.C.G.\ and L.M.C.\ contributed equally.}
\affiliation{Catalan Institute of Nanoscience and Nanotechnology (ICN2), CSIC and BIST, Campus UAB, Bellaterra, 08193 Barcelona, Spain}

\author{Frederico Sim\~oes}
\affiliation{Catalan Institute of Nanoscience and Nanotechnology (ICN2), CSIC and BIST, Campus UAB, Bellaterra, 08193 Barcelona, Spain}
\affiliation{Department of Physics, Universitat Aut{\`o}noma de Barcelona (UAB), Campus UAB, Bellaterra, 08193 Barcelona, Spain}

\author{Aron W. Cummings}
\affiliation{Catalan Institute of Nanoscience and Nanotechnology (ICN2), CSIC and BIST, Campus UAB, Bellaterra, 08193 Barcelona, Spain}

\author{Yuan Ping}
\email{yping3@wisc.edu}
\affiliation{Department of Materials Science and Engineering, University of Wisconsin-Madison, 53706, United States}
\affiliation{Department of Physics, University of Wisconsin-Madison, Madison, Wisconsin 53706, United States}
\affiliation{Department of Chemistry, University of Wisconsin-Madison, Madison, Wisconsin 53706, United States}

\date{\today}

\begin{abstract}
\noindent
Orbital angular momentum (OAM) is a promising degree of freedom for low-dissipation transport and magnetization control, yet its relaxation mechanisms remain controversial, with atom-centered approximations (ACA) predicting much shorter OAM diffusion lengths than experiments.
We address this discrepancy using first-principles Lindbladian density-matrix dynamics, capturing electron-phonon scattering and itinerant OAM contributions, together with a first-principles-parameterized tight-binding approach that separates the ACA and itinerant components.
In MoS$_2$, a strong-spin-orbit-coupling (SOC) system, orbital relaxation is multi-timescale, with fast intervalley redistribution followed by slower decay coupled to the spin.
In weak-SOC silicene, spin and orbital dynamics decouple; an electric field tunes spin relaxation while leaving orbital lifetimes unchanged.
In both materials, ACA orbital lifetimes are at least one order of magnitude shorter than itinerant ones, due to ultrafast precession driven by crystal-field splitting, absent from the itinerant component.
These results demonstrate that going beyond atom-centered models is essential for describing orbital relaxation and diffusion.
\end{abstract}

\maketitle

Orbitronics seeks to exploit the orbital angular momentum (OAM) of electrons, enabling low-dissipation electronics and providing an efficient means to control magnetization~\cite{ManchonROMP2019, go2021orbitronics, jo2024spintronics, cysne2025orbitronics}.
In contrast to charge-to-spin conversion, which relies on strong spin-orbit coupling (SOC), orbital currents can be generated in weak-SOC materials with sufficient orbital hybridization~\cite{Go2018}.
Following the prediction of a ``spin Hall effect without spins''~\cite{Bernevig2005}, tight-binding (TB) linear-response and first-principles studies demonstrated sizable orbital currents can emerge independently of SOC~\cite{tanaka2008intrinsic, kontani2009giant, jo2018gigantic}, confirmed in Ti~\cite{choi2023observation} and Cr~\cite{Lyalin2023}. More recent work demonstrates that the interplay between orbital and spin degrees of freedom can enhance spin-orbit torque efficiencies~\cite{gupta2025}, allowing light transition metals such as Mo to rival W-based devices~\cite{wang2024}.
% [v6 cut] These advances highlight the potential of orbitronics for next-generation information technologies.

Despite this rapid progress, a fundamental question remains: \emph{what controls the relaxation of nonequilibrium orbital angular momentum?}
Experiments and theory provide conflicting answers.
Experimentally, orbital diffusion lengths span a wide range: magneto-optical, nonlocal transport, orbital torque, and THz emission measurements find $\sim$70--80~nm~\cite{choi2023observation, gao2025nonlocal, hayashi2023observation, seifert2023}, whereas THz emission in CuO$_x$ and orbital Hanle experiments in Mn report only 2--9~nm~\cite{xu2024orbitronics,sala2023orbital,kang2026orbital}.
Theoretical studies display an equally striking disparity.
Two-terminal first-principles quantum transport simulations based on the atom-centered approximation (ACA) predict orbital relaxation lengths $<1$ nm~\cite{rang2024orbital}, whereas nonlocal transport calculations in multi-orbital square-lattice nanoribbons predict orbital responses that remain sizable over hundreds or even thousands of lattice constants~\cite{barbosa2025extrinsic}.

The origin of this discrepancy remains unclear because existing approaches rely on different descriptions of OAM.
In the aforementioned first-principles transport calculations, OAM is projected onto atomic orbitals and lattice sites and is treated as a purely local quantity~\cite{rang2024orbital}.
Dynamical studies based on similar localized descriptions attribute orbital relaxation primarily to crystal-field effects acting on atomic orbitals~\cite{Han2022, sohn2024, manchon2025multipolar}.
In contrast, the modern theory of magnetization has established that OAM also contains itinerant contributions associated with the geometric structure of Bloch states~\cite{thonhauser2005orbital, cysne2022orbital, burgos2024orbital, lee2026anatomy, sastges2026modern}.
These contributions are absent from the ACA and localized projections of OAM.
At the same time, they may already be implicitly present in multiorbital quantum transport simulations, which infer orbital transport from the orbital character of transmitted states rather than from a direct evaluation of OAM~\cite{barbosa2025extrinsic}.
Indeed, recent studies of graphene-based devices have shown itinerant OAM can dominate nonlocal transport responses~\cite{SalvadorSanchez2024}.
Thus, the wide range of predictions for orbital relaxation may originate from the underlying definition of orbital angular momentum itself, not differences in material systems or methodologies.

Addressing this requires a framework treating local and itinerant OAM on equal
footing while incorporating quantum scattering and relaxation dynamics. Here we
combine first-principles Lindbladian density-matrix dynamics with
electron-phonon scattering~\cite{xu2021ab, xu2024spin} and a Berry-phase formulation
of OAM~\cite{grieder2025relation} that captures itinerant contributions, and compare it
with real-space tight-binding quantum transport~\cite{fan2021linear} in which OAM is
evaluated both within the ACA and including the itinerant
component~\cite{canonico2024orbital}. Applied to monolayer MoS$_2$ (strong SOC) and silicene (weak SOC) to
disentangle spin and orbital relaxation channels, both approaches show that ACA orbital lifetimes are at
least an order of magnitude shorter than itinerant ones, because fast
crystal-field precession affects only the atom-centered component. Itinerant
OAM thus emerges as the quantity that governs orbital relaxation in these systems.

\begin{comment}
Addressing this requires a framework treating local and itinerant OAM on equal footing while incorporating quantum scattering and relaxation dynamics.
We present such a framework, based on first-principles Lindbladian density-matrix dynamics, which includes electron-phonon scattering~\cite{xu2021ab, xu2024spin} and a Berry-phase formulation of OAM~\cite{grieder2025relation} capturing itinerant contributions.
We apply this to monolayer MoS$_2$ (strong-SOC) and silicene (weak-SOC), to disentangle spin and orbital relaxation channels.
To separate and directly assess the role of itinerant OAM, we compare with tight-binding linear-scaling quantum transport calculations~\cite{fan2021linear}, in which OAM is evaluated both within the conventional ACA and using a real-space representation of the itinerant component~\cite{canonico2024orbital}.
In both cases, orbital lifetimes arising from the ACA are at least one order of magnitude shorter than those predicted from the itinerant OAM.
The short ACA lifetimes arise from fast precession induced by large crystal-field splitting, which is absent from the itinerant component.
Our results establish itinerant OAM as a fundamental ingredient governing orbital relaxation, and present a unified methodological framework for understanding orbital transport across weak and strong-SOC materials.
\end{comment}

\emph{Spin and orbital lifetimes from first-principles density-matrix dynamics (FPDMD)} --
In our first-principles approach, OAM is computed from $k$-space derivatives of Bloch states within a Berry-phase formalism, consistent with the modern theory of polarization~\cite{thonhauser2005orbital, cysne2022orbital}, which includes itinerant contributions to OAM.
% [v6 cut] This allows for a unified treatment of local and nonlocal orbital character.
%, essential for an accurate description of relaxation processes.
This implementation has been benchmarked against the Landau $g$-factor~\cite{xu2024spin, quinton2025}, circular dichroism~\cite{multunas2023circular, grieder2025relation}, and CD-ARPES~\cite{grieder2025relation}.
The OAM is computed as
\begin{equation}
      \mathbf{L}_{\mathbf{k},nm} = i \left\langle \frac{\partial u_{\mathbf{k},n}}{\partial \mathbf{k}} 
      \left|\hat{H}-\frac{\epsilon_{\mathbf{k},n}+\epsilon_{\mathbf{k},m}}{2} \right|
      \frac{\partial u_{\mathbf{k},m}}{\partial \mathbf{k}}\right\rangle,
\end{equation}
where $\hat{H}$ is the Hamiltonian operator, $u_{\mathbf{k},n}$ is the periodic part of the Bloch wavefunction, $n$ and $m$ are band indices, and $\epsilon$ is the corresponding eigenenergy.

We employ a first-principles technique based on Lindbladian dynamics of the one-particle electron density matrix with explicit electron-phonon scattering and spin-orbit coupling~\cite{xu2021ab, xu2023ab}, which accurately predicts spin relaxation and dephasing times in disparate solids~\cite{xu2024spin, xu2021giant, xu2020spin} and was recently generalized to non-Markovian dynamics of coupled electron-phonon systems~\cite{simoni2025, riva2026}.
The equation of motion is
\begin{equation}
	\frac{d\rho_{12}}{dt} = \frac{1}{2} \sum_{345} 
	\left[ 
	\begin{array}{lr}
		(I-\rho)_{13}P^\mathrm{e-ph}_{32,45}\rho_{45} \\
		-(I-\rho)_{45}P^\mathrm{e-ph,*}_{45,13}\rho_{32}
	\end{array}
	\right] + \mathrm{H.c.},
\end{equation}
where $\rho$ is the one-particle density matrix built from Kohn-Sham states, the indices ($1-5$) represent a combination of band and $k$-point indices, and $P^\mathrm{e-ph}$ is the electron-phonon scattering matrix, constructed from first-principles electron-phonon matrix elements~\cite{xu2020spin} as detailed in Supplemental Material (SM) Sec.~VII~\cite{suppmaterial}.
Observables for spin and OAM relaxation are computed as
\begin{equation}
	\mathcal{O}_i(t) = \mathbf{Tr}[\mathcal{O}_i \rho(t)],
\end{equation}
where $\mathcal{O} = S,L$ for spin or OAM, respectively, and $i$ is the Cartesian direction.
Spin and orbital lifetimes ($\tau_{\mathcal{O}_i}$) are computed by fitting $S_i(t)$ or $L_i(t)$ to an exponential decay.

\emph{Real-space spin and orbital dynamics} --
Spin and orbital relaxation dynamics are computed in real space using a linear-scaling method based on Chebyshev polynomial expansion of spectral functions~\cite{fan2021linear, joao2020kite}. Because this approach employs TB Hamiltonians, the atom-centered and itinerant components of the OAM can be separated straightforwardly.
Observables are computed as
\begin{equation}
\langle \mathcal{O}_i(t,E) \rangle =
\frac{
\langle \psi_{P_{\mathcal{O}_i}} |
\hat{U}^{\dagger}(t) \mathcal{O}_i \delta(\hat{H}-E) \hat{U}(t) |
\psi_{P_{\mathcal{O}_i}} \rangle
}
{
\langle \psi_{P_{\mathcal{O}_i}} |
\delta(\hat{H}-E) |
\psi_{P_{\mathcal{O}_i}} \rangle
},
\end{equation}
where $\hat{U}$ is the time evolution operator and $|\psi_{P_{\mathcal{O}_i}}\rangle = P_{\mathcal{O}_i}|\psi\rangle$ is a wavepacket polarized with respect to $\mathcal{O}_i$, which corresponds to the spin operator $S_i$, the atom-centered orbital angular momentum $L_i^\mathrm{ACA}$, or the full OAM including itinerant contributions $L_i^\mathrm{Berry}$.
To relate spin and orbital dynamics to charge transport, the momentum relaxation time $\tau_p$ is extracted from the long-time diffusive regime of the wavepacket mean-squared displacement (SM Secs.~I and II~\cite{suppmaterial}).

\begin{figure}[!ht]
	\centering
	\includegraphics[width=0.95\linewidth]
	{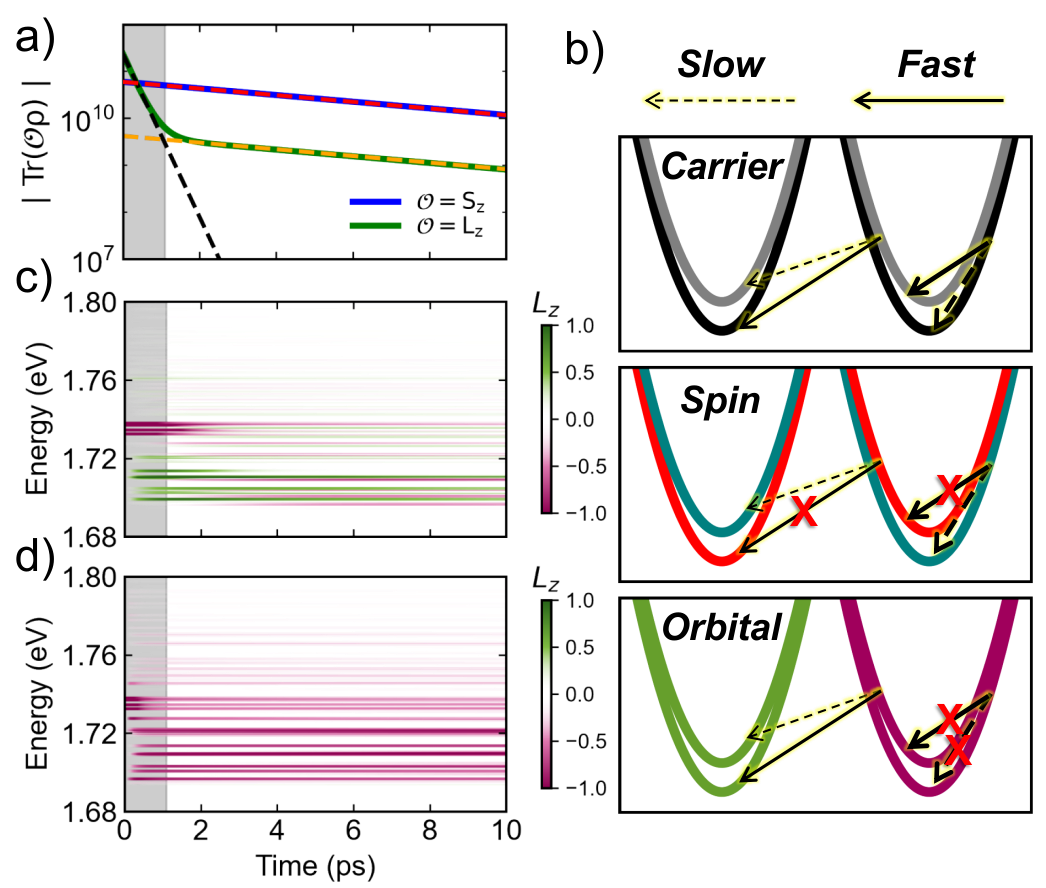}
	\caption{
		Spin and orbital dynamics of optically excited electrons  in MoS$_2$. (a) Spin and orbital relaxation at 300 K. Dashed lines indicate the two decay time scales, 0.25 ps (black) and 6.1 ps (orange, red). (b) Schematics of carrier, spin, and orbital inter- and intra-valley relaxation pathways. (c) Energy-resolved orbital relaxation for only intervalley relaxation, and (d) intravalley relaxation. The color scale indicates normalized orbital polarization.
	}
	\label{fig:MoS2-relax}
\end{figure}

\emph{OAM relaxation in the strong SOC regime (MoS$_2$) --}
First we investigate spin and orbital relaxation in monolayer MoS$_2$, a prototypical system with strong SOC and valley-dependent orbital texture.
The system is initialized with a circularly-polarized pump with an excitation energy just above the band gap at 1.7 eV, and the resulting dynamics of the out-of-plane (along $z$) spin and OAM are shown in Fig.~\ref{fig:MoS2-relax}.

Orbital relaxation exhibits two distinct timescales.
The initial relaxation, in the gray region in Fig.~\ref{fig:MoS2-relax}, is similar to the carrier relaxation time ($0.05$ ps) with a fast orbital decay (lifetime of $0.25$ ps), absent from the spin dynamics.
At longer times the spin and orbital polarization both decay with a lifetime of $6.1$ ps.

The initial fast orbital decay originates from intervalley scattering.
The circularly-polarized pump populates a single valley, producing a larger orbital than spin polarization at $t=0$ (Fig.~\ref{fig:MoS2-relax}(a)).
Carriers then rapidly scatter into both valleys, occupying states of opposite orbital but identical spin polarization (Fig.~\ref{fig:MoS2-relax}(b)), and the net OAM therefore drops while the spin does not.
This is confirmed by restricting the scattering channels: with only intervalley scattering (Fig.~\ref{fig:MoS2-relax}(c)), the initially all-positive orbital polarization redistributes toward a mix of positive and negative states before relaxing in step with the spin, whereas with only intravalley scattering (Fig.~\ref{fig:MoS2-relax}(d)) the population redistributes in energy but, confined to one valley, retains its orbital polarization sign. A magnetic-field initialization with no hot carriers yields the same two timescales (SM Sec.~VII D), ruling out carrier cooling as the origin of the fast component.

This highlights a key distinction between spin and orbital dynamics: \textit{the total OAM includes a component tied to the valley polarization, and one tied to the spin through the strong SOC.}
The initial orbital decay reflects redistribution across valleys and requires no spin-flip processes.
The long-time dynamics, in contrast, is governed by strong SOC, which locks the spin and orbital degrees of freedom into a common relaxation time.

Next we use the real-space TB approach to examine which component of OAM drives the relaxation process.
We describe MoS$_2$ with a pseudo-atomic orbital (PAO) TB Hamiltonian~\cite{suppmaterial} constructed from first principles using PAOFLOW~\cite{PAO6}, with Anderson disorder tuned to reproduce the FPDMD spin lifetime of Fig.~\ref{fig:MoS2-relax}.
The ACA contribution alone gives an orbital lifetime of $<3$~fs, three to four orders of magnitude shorter than the FPDMD result, which uses the full Berry-phase OAM comprising both ACA and itinerant contributions. Orbital relaxation in MoS$_2$ is therefore governed by itinerant OAM, the contribution central to the modern theory of orbital magnetization, which the ACA omits by construction.
We quantify this contrast further for silicene below.

\textit{OAM relaxation in the weak SOC regime (silicene)} --
Next we consider silicene, whose weak SOC isolates the orbital relaxation mechanism, a direct counterpoint to MoS$_2$.
As before, FPDMD yields the intrinsic, electron-phonon-limited orbital lifetime.
In parallel, our real-space approach employs a multi-orbital TB model with $s$ and $p$ orbitals, atomic SOC, and a Stark term for the out-of-plane electric field, parametrized to the DFT band structure~\cite{suppmaterial}.
Crucially, this model retains both the atom-centered and itinerant OAM on equal footing: the former
arises from hybridization between the $p_z$ and the $s$, $p_x$, and $p_y$ orbitals of the buckled lattice, while the latter originates from the self-rotation of Bloch wave packets near the $K/K'$ valleys~\cite{Xiao2007, Bhowal2021, pezo2023orbital}.

\begin{figure}[!ht]
	\centering
	\includegraphics[width=0.9\linewidth]{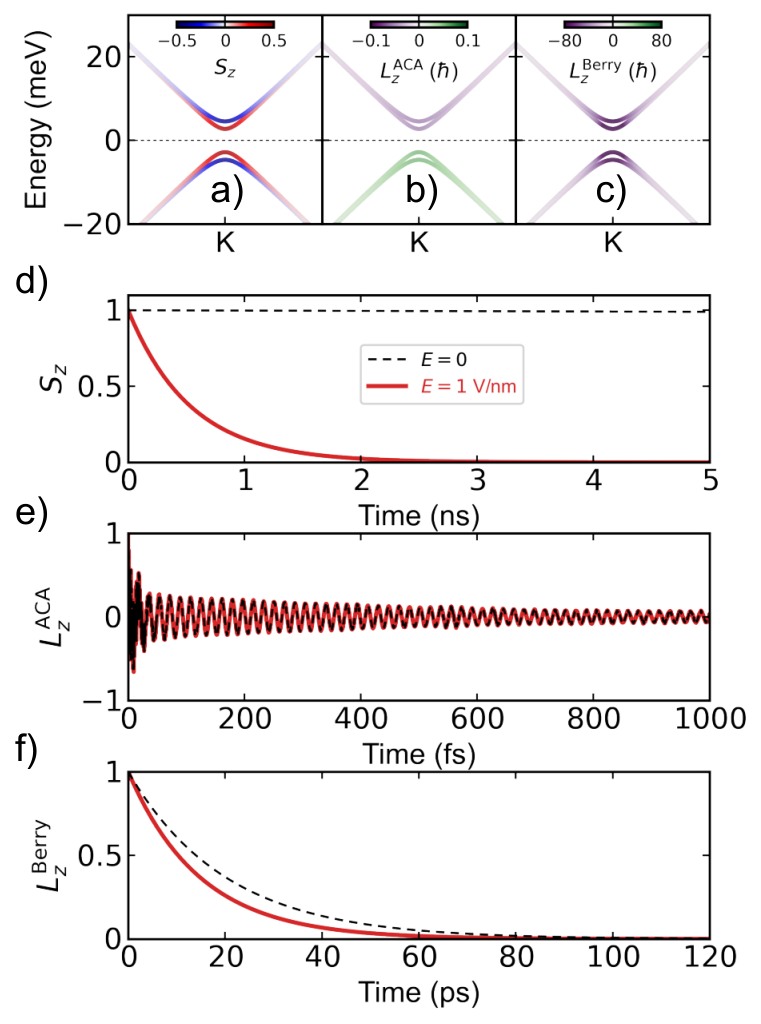}
	\caption{
		Spin and orbital textures and dynamics in silicene near the $K$ valley. (a-c) Band structure at $E=1$~V/nm with color indicating (a) spin, (b) atom-centered OAM, and (c) itinerant OAM. Note the opposite (same) sign of $S_z$ ($L_z$) of the field-split branches. (d-f) Polarization dynamics at $E=0$ (dashed) and $E=1$~V/nm (solid) %\textcolor{red}{at 100K} 
        for (d) spin by FPDMD at $100$~K, (e) atom-centered OAM by TB, and (f) itinerant OAM by FPDMD at $100$~K.}
        
	\label{fig:silicene-bands}
\end{figure}

Figure~\ref{fig:silicene-bands} contrasts the equilibrium textures and the nonequilibrium dynamics of the spin and of the two formulations of OAM.
The spin behaves as dictated by inversion and time-reversal symmetries, with vanishing texture at zero field and the field-split branches carrying \textit{opposite} $S_z$ (Fig.~\ref{fig:silicene-bands}(a)).
The orbital textures behave differently, with the field-split branches carrying the \emph{same} orbital polarization (Fig.~\ref{fig:silicene-bands}(b,c)), indicating inversion-symmetry breaking plays a fundamentally different role for OAM.
The two orbital formulations also differ: within the ACA the conduction and valence bands carry small ($< 0.1 \hbar$) orbital moments of opposite sign, whereas the itinerant OAM has the same sign in both bands and reaches values nearly three orders of magnitude larger (up to $80\hbar$)~\cite{lee2026anatomy}.

The scattering-driven spin and orbital dynamics, in Fig.~\ref{fig:silicene-bands}(d-f), reveal additional contrasting behavior.
Spin relaxation is enhanced dramatically between $E=0$ and finite field, driven by field-induced spin splitting and precession (Fig.~\ref{fig:silicene-bands}(d)).
In sharp contrast, the ACA and Berry-phase orbital dynamics are nearly indistinguishable at $E=0$ and $1$~V/nm (Fig.~\ref{fig:silicene-bands}(e,f)).
The ACA component oscillates rapidly with a frequency $\omega_{{p_z}} \approx 0.7\times \Delta_{{p_z}} / \hbar$, where $\Delta_{{p_z}} \approx 0.52$~eV  is the energy difference of the $p_{x,y}$ and $p_z$ orbitals and decays on the sub-ps time scale.
Meanwhile, the itinerant OAM exhibits little oscillation and decays on a timescale of tens of picoseconds, much slower than the ACA OAM.

\begin{figure}[htbp!]
	\centering
	\includegraphics[width=\linewidth]{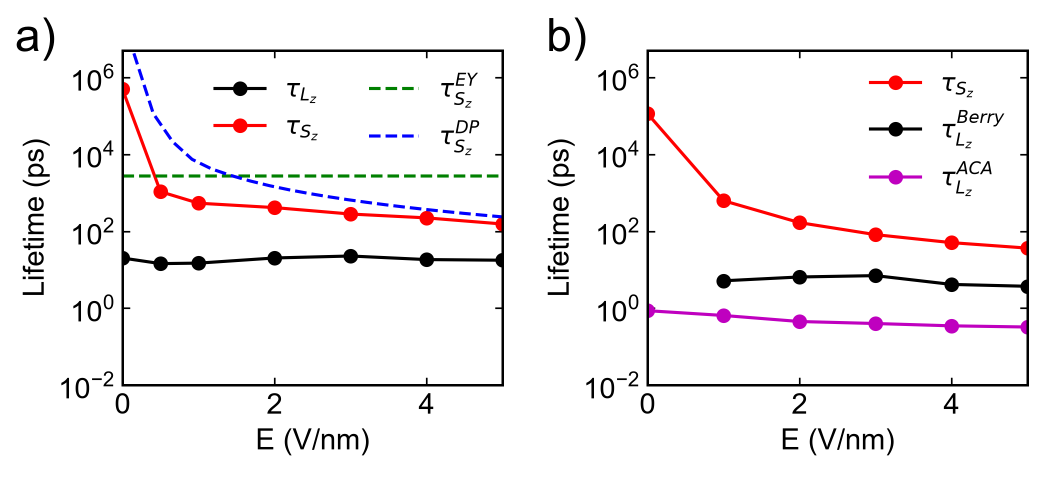}
	\caption{
		Spin and orbital lifetimes in silicene, with the Fermi level at the conduction band minimum, as a function of applied electric field. (a) Lifetimes arising from FPDMD with electron-phonon scattering at $100$~K, and (b) lifetimes from real-space TB dynamics with Anderson disorder strength $W=700$~meV. 
        In (a), $\tau^\mathrm{EY}_{S_z}$ and $\tau^\mathrm{DP}_{S_z}$ are evaluated from the EY and DP models of spin relaxation. In (b), $\tau^\mathrm{Berry}_{L_z}$ is the orbital lifetime including the itinerant contributions and $\tau^\mathrm{ACA}_{L_z}$ only considers the atom-centered component.
	}
	\label{fig:silicene-relax-Efield}
\end{figure}

Figure~\ref{fig:silicene-relax-Efield} shows the spin and orbital lifetimes as a function of electric field, calculated from (a) FPDMD and (b) real-space TB dynamics.
The spin lifetimes from the two approaches show the same electric field dependence, consistent with mixed Elliott-Yafet (EY) and D'yakonov-Perel' (DP) relaxation already established for silicene in Ref.~\cite{xu2021giant}, validating the TB parametrization.
More importantly, the first-principles orbital lifetime matches the itinerant (Berry) OAM in the TB dynamics, despite the different scattering sources (electron-phonon vs.\ Anderson disorder).
%Both reveal orbital lifetimes on the order of tens of ps. 
%
Both agree to within a factor of $\sim$3, on the $\sim$10 ps scale.
In sharp contrast, the ACA orbital polarization decays biexponentially.
The faster component, shorter than the momentum relaxation time, arises from dephasing of the coherent crystal-field precession rather than from scattering; we therefore identify the ACA orbital lifetime with the slower, scattering-limited component.
Even this lifetime is one order of magnitude shorter than the itinerant one at all fields, on the order of $1$~ps.
As already indicated in Fig.~\ref{fig:silicene-bands}, both orbital lifetimes are nearly independent of the field. 
%
%\revb{Notably, the FPDMD orbital lifetime is unchanged down to zero field, and the ACA precession scale $\Delta_{p_z}$ is set by the lattice rather than the electric field~\cite{Han2022, sohn2024, manchon2025multipolar}.
%This shows that inversion-symmetry breaking, which drives the spin relaxation here, is not required for orbital relaxation.}
%\textcolor{red}{[This argument was already made in a couple different places below, after Fig.~3. Is it premature to include this here? Also for space considerations...]}
%\textcolor{red}{[Luis/Andrew: please confirm from your calculations the zero-field intervalley picture and whether the $E=0$ decay can be equivalently read as valley relaxation observed through the OAM.]}
Together, these results establish the itinerant component as the dominant channel for orbital dynamics in multiorbital Dirac materials.

The OAM dynamics, and their apparent independence from electric field, suggests each component is controlled by an intrinsic scale of the material.
For ACA OAM, precession is driven by the crystal field splitting, as shown in Fig.~\ref{fig:silicene-bands}(e), which is independent of the external field and requires no symmetry breaking.
This is consistent with previous work on orbital dynamics and relaxation~\cite{Han2022, sohn2024, manchon2025multipolar},  establishing ACA OAM dynamics as an intrinsic property of the lattice.
Meanwhile, for the itinerant OAM, each Bloch eigenstate carries a finite moment, $\mathbf{L}_{\mathbf{k},nn} \neq 0$, while the off-diagonal elements, $\mathbf{L}_{\mathbf{k},n \neq m}$, are an order of magnitude smaller near the band edge in the first-principles calculations (see SM Fig.~S9)~\cite{suppmaterial}.
As such, the itinerant OAM exhibits minimal precession, and must therefore relax through scattering that connects states of opposite OAM in $k$-space.
In silicene, this relaxation channel is set by the valley separation and the associated intervalley scattering time, again independent of electric field.
This suggests that, in general, the relaxation of itinerant OAM is set by the distribution of its texture in $k$-space, as examined in Fig.~\ref{fig:silicene-intervalley} below.

\textit{Analysis of relaxation mechanisms} -- 
We examine the nature of spin and orbital relaxation by calculating their scaling with the momentum relaxation time, $\tau_p$, which is done in the FPDMD and TB calculations by scaling the scattering matrix and by varying the Anderson disorder strength, respectively.
Spin relaxation is conventionally classified by the EY mechanism, $\tau_s \propto \tau_p$, or the DP mechanism, $\tau_s \propto 1/\tau_p$~\cite{zutic2004}.
Both numerical approaches reproduce the known spin relaxation behavior of silicene -- EY relaxation at $E=0$ with a crossover to mixed EY+DP at finite field~\cite{xu2021giant} -- as shown in SM Sec.~VI~\cite{suppmaterial}.

\begin{figure}[ht]
	\centering
	\includegraphics[width=\linewidth]{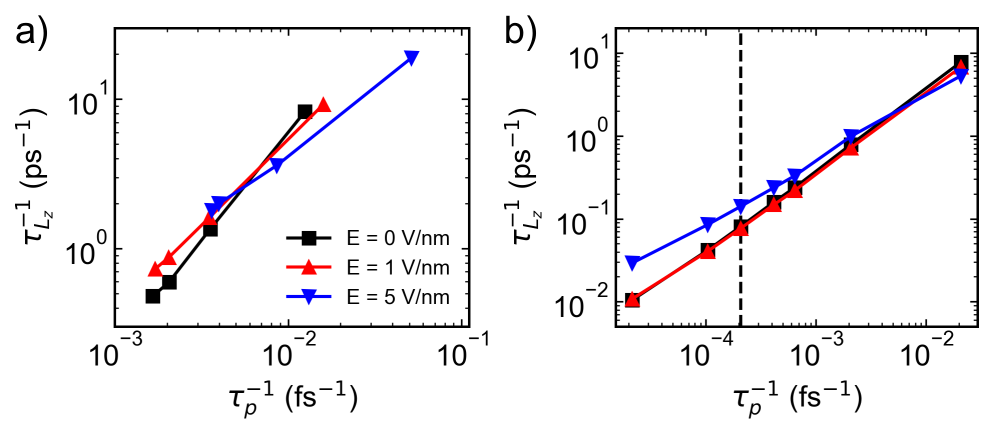}
	\caption{
        Orbital relaxation rate as a function of momentum relaxation rate, for (a) the TB-ACA and (b) the FPDMD (Berry-phase) OAM at 100~K. Colors denote different applied electric fields. The dashed line in (b) denotes the intrinsic electron-phonon scattering rate.
    }
	\label{fig:silicene-relax-mech}
\end{figure}

The scaling of orbital relaxation rate with $\tau_p^{-1}$ is shown in Fig.~\ref{fig:silicene-relax-mech}.
For the atom-centered component, in panel (a), %scales as $\tau_{L_z}^{-1} \propto \tau_p^{-1}$ at all fields.
$\tau_{L_z}^{-1}$ scales approximately linearly with $\tau_p^{-1}$ at all fields, with a slightly weaker dependence at 5 V/nm.
Although this mimics EY, the mechanism is distinct.
In the strong-precession regime set by the crystal field, $\omega_{p_z} \tau_p \gg 1$, rapid orbital precession is interrupted by charge scattering, yielding $\tau_{L_z} \approx \tau_p$.
This is analogous to the strong precession regime of spin relaxation, where the spin lifetime likewise becomes approximately equal to the momentum relaxation time~\cite{Gridnev2001, Brand2002}, and has also been predicted in prior studies of ACA orbital dynamics~\cite{manchon2025multipolar}.
Because $\Delta_{p_z}$ is set by the lattice, this channel is largely field-independent.
% [v6] removed a stray "\\" line break here (produced an empty line / underfull hbox).
%\textcolor{red}{[We see similar behavior for both ACA and itinerant -- a move to more DP-like at high field. We emphasize this transition in the itinerant component (Fig.~5), but ignore it for the ACA.]}
%\\
%\textcolor{red}{[Also, for the same scattering rates we now see the same lifetime for both ACA and itinerant -- see the example I sent vial email.]}

The itinerant component, in Fig.~\ref{fig:silicene-relax-mech}(b), is EY-like at low fields and acquires a slightly mixed EY+DP character at $5$~V/nm, which we examine via the valley-resolved decomposition in Fig.~\ref{fig:silicene-intervalley}.
Intervalley scattering dominates itinerant orbital relaxation at all fields (panel (a)), but its share decreases with increasing field (panel (b)) because the field enhances the in-plane orbital texture (insets), opening an intravalley relaxation channel.
The growth of intravalley relaxation offsets the weakening intervalley contribution, leaving the total lifetime nearly field-independent while slightly altering the overall scaling with $\tau_p$.
This behavior demonstrates that relaxation of itinerant OAM is determined by the distribution of orbital texture in momentum space, in contrast to the ACA OAM, whose relaxation is set by crystal-field-driven precession.

\begin{figure}[ht]
	\centering
	\includegraphics[scale=0.43, trim=4 4 4 4,clip]{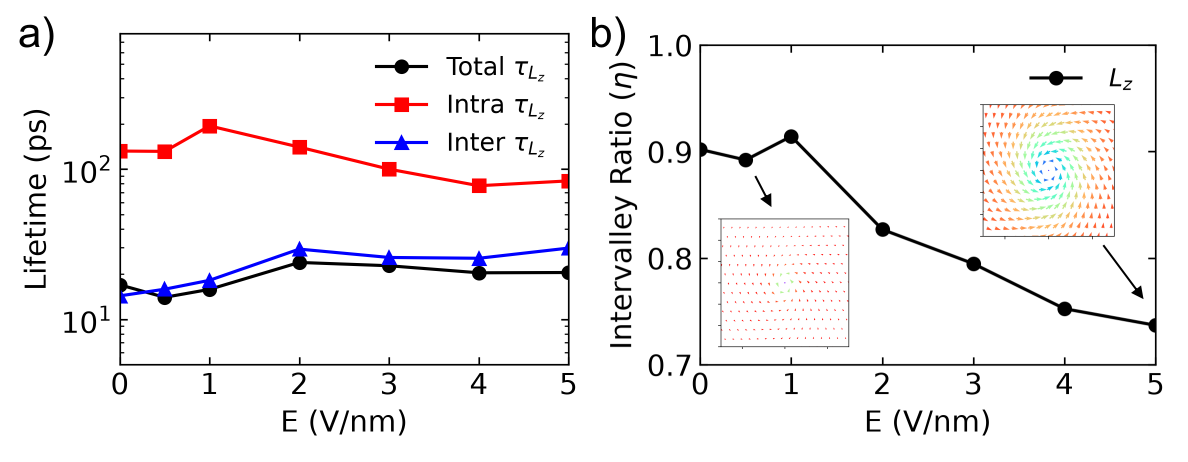}
	\caption
	{
		Orbital lifetime by FPDMD partitioned into scattering channels as a function of applied electric field. (a) Contribution from only intravalley scattering (Intra $\tau_{L_z}$), intervalley scattering (Inter $\tau_{L_z}$), and both channels (Total $\tau_{L_z}$). (b) The proportion of relaxation coming from the intervalley channel, $\eta = (\tau^\mathrm{inter}_{L,i} ) ^{-1}/[(\tau^\mathrm{inter})^{-1}+w(\tau^\mathrm{intra}_{L,i})^{-1}]$~\cite{suppmaterial}, with insets depicting the orbital texture at different applied fields.
    }
	\label{fig:silicene-intervalley}
\end{figure}

\emph{Summary and conclusions} --
We have developed a first-principles framework based on Lindbladian density-matrix dynamics for OAM relaxation which unifies local orbital character, itinerant Berry-phase contributions, and quantum electron-phonon scattering, independently cross-validated against a real-space linear-scaling TB transport approach.
The two methods yield a consistent microscopic picture.
For strong SOC (MoS$_2$), the nonequilibrium orbital polarization decays on two distinct timescales: a fast intervalley redistribution of the valley-polarized population, followed by intrinsic relaxation locked to the spin by SOC.
In the weak-SOC limit (silicene), spin and orbital dynamics decouple, with the spin lifetime strongly modulated by an electric field while orbital relaxation remains field-independent.

Most importantly, we find distinct relaxation mechanisms, and corresponding time scales, for the ACA and itinerant components of OAM.
The ACA component exhibits fast precession, driven by crystal field splitting, and relaxes on short timescales of the order of the charge scattering time.
Meanwhile, itinerant OAM is carried by Bloch states, and relaxation is determined by its distribution of orbital texture in momentum space.
In silicene, itinerant orbital lifetime is tied to intervalley scattering, which connects pockets of opposite OAM in opposite valleys.
Here, orbital dynamics is dominated by itinerant OAM, leading to orbital lifetimes that are one order of magnitude longer than a purely atom-centered approach.
Theoretical studies of orbital dynamics, and the interpretation of experiments, should therefore include the itinerant component of OAM---certainly in MoS$_2$ and silicene, and likely across the larger family of 2D materials.

In metals, we expect the balance of the two channels to depend on both orbital character and sample quality.
The atomic contribution dominates the orbital magnetization of localized $3d$ metals but fails for delocalized $sp$ and $5d$ metals~\cite{lee2026anatomy}, while the crystal-field channel crosses from EY-like relaxation in clean crystals to motionally-narrowed, DP-like relaxation in the strongly disordered limit~\cite{sohn2024, manchon2025multipolar}.
This crossover has recently been observed in Mn films with variable disorder, where the orbital lifetime \emph{decreases} with increasing crystalline order and transitions to EY-like in single-crystalline $\alpha$-Mn~\cite{kang2026orbital}.
The wide spread of measured orbital diffusion lengths~\cite{choi2023observation, seifert2023, xu2024orbitronics, sala2023orbital} may partly reflect this interplay.

\begin{acknowledgments} 
\textbf{Acknowledgments} --
This work is primarily supported by the Computational Chemical Sciences program within the Office of Science of the DOE under Grant No.\ DE-SC0023301 for the code development, and partially supported by the support for materials theory application as part of the Center for Hybrid Organic-Inorganic Semiconductors for Energy (CHOISE), an Energy Frontier Research Center funded by the Office of Basic Energy Sciences, Office of Science within the US Department of Energy (DOE). 
Calculations were carried out at the National Energy Research Scientific Computing Center (NERSC), a U.S.\ Department of Energy Office of Science User Facility operated under Contract No.\ DEAC02-05CH11231.
This work used the TACC Stampede3 system at the University of Texas at Austin through allocation PHY240212 from the Advanced Cyberinfrastructure Coordination Ecosystem: Services and Support (ACCESS) program~\cite{Boerner2023}, which is supported by U.S.\ National Science Foundation grant Nos.\ 2138259, 2138286, 2138307, 2137603, and 2138296. L.M.C., F.S., and A.W.C.\ thankfully acknowledge the computer resources at MareNostrum 5 and the technical support provided by the Barcelona Supercomputing Center (BSC), through the Red Española de Supercomputación (RES), under activity ID FI-2026-1-0042. ICN2 is funded by the CERCA programme / Generalitat de Catalunya, and is supported by the Severo Ochoa Centres of Excellence programme, Grant CEX2021-001214-S, funded by MCIU/AEI/10.13039.501100011033. The project that gave rise to these results received the support of a fellowship from the ``la Caixa'' Foundation (ID 100010434). The fellowship code is LCF/BQ/DFI25/13000084 (F.S.).
\end{acknowledgments}

\bibliography{ref}

@article{kang2026orbital,
  title   = {Orbital Hall conductivity and relaxation in thin films with variable disorder},
  author  = {Kang, Min-Gu and Nasr, Federica and Sala, Giacomo and Schlitz, Richard and Ding, Shilei and Bartsch, Stefan and Rossell, Marta D. and Alvarado, Santos F. and Gambardella, Pietro},
  journal = {Nature Physics},
  year    = {2026},
  doi     = {10.1038/s41567-026-03334-z}
}

@article{zutic2004,
  title     = {Spintronics: Fundamentals and applications},
  author    = {{\v{Z}}uti{\'c}, Igor and Fabian, Jaroslav and Das Sarma, Sankar},
  journal   = {Reviews of Modern Physics},
  volume    = {76},
  issue     = {2},
  pages     = {323--410},
  year      = {2004},
  publisher = {American Physical Society},
  doi       = {10.1103/RevModPhys.76.323}
}

@inproceedings{Boerner2023,
author = {Boerner, Timothy J. and Deems, Stephen and Furlani, Thomas R. and Knuth, Shelley L. and Towns, John},
title = {ACCESS: Advancing Innovation: NSF’s Advanced Cyberinfrastructure Coordination Ecosystem: Services \& Support},
year = {2023},
isbn = {9781450399852},
pubsher = {Association for Computing Machinery},
address = {New York, NY, USA},
url = {https://doi.org/10.1145/3569951.3597559},
doi = {10.1145/3569951.3597559},
booktitle = {Practice and Experience in Advanced Research Computing 2023: Computing for the Common Good},
pages = {173–176},
numpages = {4},
location = {Portland, OR, USA},
series = {PEARC '23}
}

@misc{riva2026,
      title={Open-quantum-system theory of non-Markovian electron-phonon dynamics}, 
      author={Gabriele Riva and Jacopo Simoni and Yuan Ping},
      year={2026},
      eprint={2606.22233},
      archivePrefix={arXiv},
      primaryClass={cond-mat.mtrl-sci},
      url={https://arxiv.org/abs/2606.22233}, 
}

@article{joao2020kite,
  title={KITE: high-performance accurate modelling of electronic structure and response functions of large molecules, disordered crystals and heterostructures},
  author={Jo{\~a}o, Sim{\~a}o M and An{\dj}elkovi{\'c}, Mi{\v{s}}a and Covaci, Lucian and Rappoport, Tatiana G and Lopes, Jo{\~a}o MVP and Ferreira, Aires},
  journal={Royal Society Open Science},
  volume={7},
  number={2},
  pages={191809},
  year={2020}
}

@article{Bhowal2021,
  title = {Orbital Hall effect as an alternative to valley Hall effect in gapped graphene},
  author = {Bhowal, Sayantika and Vignale, Giovanni},
  journal = {Phys. Rev. B},
  volume = {103},
  issue = {19},
  pages = {195309},
  numpages = {8},
  year = {2021},
  month = {May},
  publisher = {American Physical Society},
  doi = {10.1103/PhysRevB.103.195309},
  url = {https://link.aps.org/doi/10.1103/PhysRevB.103.195309}
}

@article{PAO6,
title = {Advanced modeling of materials with PAOFLOW 2.0: New features and software design},
journal = {Computational Materials Science},
volume = {200},
pages = {110828},
year = {2021},
issn = {0927-0256},
doi = {https://doi.org/10.1016/j.commatsci.2021.110828},
url = {https://www.sciencedirect.com/science/article/pii/S0927025621005486},
author = {Frank T. Cerasoli and Andrew R. Supka and Anooja Jayaraj and Marcio Costa and Ilaria Siloi and Jagoda Sławińska and Stefano Curtarolo and Marco Fornari and Davide Ceresoli and Marco {Buongiorno Nardelli}}
}

@misc{suppmaterial,
note = {See the Supplemental Material at [URL will be inserted by the publisher] for a detailed description of the methods used to determine the real-space spin and orbital relaxation times, computational details of the first-principles parametrization of the real-space MoS$_2$ Hamiltonian, the spin and orbital lifetimes obtained within the ACA, details of the parameter adjustments and the orbital textures of silicene, and computational details of the first-principles density-matrix dynamics method.}
}

@article{Xiao2007,
  title = {Valley-Contrasting Physics in Graphene: Magnetic Moment and Topological Transport},
  author = {Xiao, Di and Yao, Wang and Niu, Qian},
  journal = {Phys. Rev. Lett.},
  volume = {99},
  issue = {23},
  pages = {236809},
  numpages = {4},
  year = {2007},
  month = {Dec},
  publisher = {American Physical Society},
  doi = {10.1103/PhysRevLett.99.236809},
  url = {https://link.aps.org/doi/10.1103/PhysRevLett.99.236809}
}

@article{burgos2024orbital,
  title={Orbital angular momentum of Bloch electrons: equilibrium formulation, magneto-electric phenomena, and the orbital Hall effect},
  author={Burgos Atencia, Rhonald and Agarwal, Amit and Culcer, Dimitrie},
  journal={Advances in Physics: X},
  volume={9},
  number={1},
  pages={2371972},
  year={2024},
  publisher={Taylor \& Francis}
}

@article{quinton2025,
  title = {Magnetic-field dependence of spin-phonon relaxation and dephasing due to $g$-factor fluctuations from first principles},
  author = {Quinton, Joshua and Fadel, Mayada and Xu, Junqing and Habib, Adela and Chandra, Mani and Ping, Yuan and Sundararaman, Ravishankar},
  journal = {Phys. Rev. B},
  volume = {111},
  issue = {11},
  pages = {115113},
  numpages = {8},
  year = {2025},
  month = {Mar},
  publisher = {American Physical Society},
  doi = {10.1103/PhysRevB.111.115113},
  url = {https://link.aps.org/doi/10.1103/PhysRevB.111.115113}
}

@article{sastges2026modern,
  title={Modern Approach to Orbital Hall Effect Based on Wannier Picture of Solids},
  author={Sastges, Mirco and Baek, Insu and Lee, Hojun and Lee, Hyun-Woo and Mokrousov, Yuriy and Go, Dongwook},
  journal={arXiv preprint arXiv:2604.08280},
  year={2026}
}

@article{lee2026anatomy,
  title={Anatomy of the modern theory of orbital magnetism from first-principles: term-by-term analysis in the gauge-covariant formalism},
  author={Lee, Hojun and Baek, Insu and Sastges, Mirco and Mokrousov, Yuriy and Lee, Hyun-Woo and Go, Dongwook},
  journal={arXiv preprint arXiv:2603.19875},
  year={2026}
}

@article{canonico2024orbital,
  title={Orbital Hall responses in disordered topological materials},
  author={Canonico, Luis M and Garcia, Jose H and Roche, Stephan},
  journal={Physical Review B},
  volume={110},
  number={14},
  pages={L140201},
  year={2024},
  publisher={APS}
}

@article{fan2021linear,
  title={Linear scaling quantum transport methodologies},
  author={Fan, Zheyong and Garcia, Jose H and Cummings, Aron W and Barrios-Vargas, Jose Eduardo and Panhans, Michel and Harju, Ari and Ortmann, Frank and Roche, Stephan},
  journal={Physics Reports},
  volume={903},
  pages={1--69},
  year={2021},
  publisher={Elsevier}
}

@article{xu2024orbitronics,
  title={Orbitronics: light-induced orbital currents in Ni studied by terahertz emission experiments},
  author={Xu, Yong and Zhang, Fan and Fert, Albert and Jaffres, Henri-Yves and Liu, Yongshan and Xu, Renyou and Jiang, Yuhao and Cheng, Houyi and Zhao, Weisheng},
  journal={Nature Communications},
  volume={15},
  number={1},
  pages={2043},
  year={2024},
  publisher={Nature Publishing Group UK London}
}

@article{SalvadorSanchez2024,
  title = {Generation and control of nonlocal chiral currents in graphene superlattices by orbital Hall effect},
  author = {Salvador-S\'anchez, Juan and Canonico, Luis M. and P\'erez-Rodr\'{\i}guez, Ana and Cysne, Tarik P. and Baba, Yuriko and Cleric\`o, Vito and Vila, Marc and Vaquero, Daniel and Delgado-Notario, Juan Antonio and Caridad, Jos\'e M. and Watanabe, Kenji and Taniguchi, Takashi and Molina, Rafael A. and Dom\'{\i}nguez-Adame, Francisco and Roche, Stephan and Diez, Enrique and Rappoport, Tatiana G. and Amado, Mario},
  journal = {Phys. Rev. Res.},
  volume = {6},
  issue = {2},
  pages = {023212},
  numpages = {11},
  year = {2024},
  month = {May},
  publisher = {American Physical Society},
  doi = {10.1103/PhysRevResearch.6.023212},
  url = {https://link.aps.org/doi/10.1103/PhysRevResearch.6.023212}
}

@article{manchon2025multipolar,
  title = {Multipolar Orbital Relaxation of the ${t}_{2g}$ States},
  author = {Manchon, Aur\'elien and Sun, Chi and Ning, Xiaobai and Sato, Tetsuya and Kato, Takeo and Rappoport, Tatiana G.},
  journal = {Phys. Rev. Lett.},
  volume = {136},
  issue = {22},
  pages = {226801},
  numpages = {8},
  year = {2026},
  month = {Jun},
  publisher = {American Physical Society},
  doi = {10.1103/h17x-qg4y},
  url = {https://link.aps.org/doi/10.1103/h17x-qg4y}
}

@article{sohn2024,
  title={Dyakonov-Perel-like orbital and spin relaxations in centrosymmetric systems},
  author={Sohn, Jeonghun and Lee, Jongjun M and Lee, Hyun-Woo},
  journal={Physical Review Letters},
  volume={132},
  number={24},
  pages={246301},
  year={2024},
  publisher={APS}
}

@article{Han2022,
  title = {Orbital Dynamics in Centrosymmetric Systems},
  author = {Han, Seungyun and Lee, Hyun-Woo and Kim, Kyoung-Whan},
  journal = {Phys. Rev. Lett.},
  volume = {128},
  issue = {17},
  pages = {176601},
  numpages = {7},
  year = {2022},
  month = {Apr},
  publisher = {American Physical Society},
  doi = {10.1103/PhysRevLett.128.176601},
  url = {https://link.aps.org/doi/10.1103/PhysRevLett.128.176601}
}

@article{barbosa2025extrinsic,
  title={Extrinsic orbital Hall effect and orbital relaxation in mesoscopic devices},
  author={Barbosa, Anderson LR and Lee, Hyun-Woo and Rappoport, Tatiana G},
  journal={arXiv preprint arXiv:2507.01941},
  year={2025}
}

@article{sala2023orbital,
  title={Orbital Hanle magnetoresistance in a 3 d transition metal},
  author={Sala, Giacomo and Wang, Hanchen and Legrand, William and Gambardella, Pietro},
  journal={Physical Review Letters},
  volume={131},
  number={15},
  pages={156703},
  year={2023},
  publisher={APS}
}

@article{seifert2023,
  title={Time-domain observation of ballistic orbital-angular-momentum currents with giant relaxation length in tungsten},
  author={Seifert, Tom S and Go, Dongwook and Hayashi, Hiroki and Rouzegar, Reza and Freimuth, Frank and Ando, Kazuya and Mokrousov, Yuriy and Kampfrath, Tobias},
  journal={Nature nanotechnology},
  volume={18},
  number={10},
  pages={1132--1138},
  year={2023},
  publisher={Nature Publishing Group UK London}
}

@article{wang2024,
  title={Large field-like spin--orbit torque and enhanced magnetization switching efficiency utilizing amorphous Mo},
  author={Wang, Xinran and Meng, Ao and Yao, Yuxuan and Lin, Fangye and Bai, Yue and Ning, Xiaobai and Li, Bo and Zhang, Yue and Nie, Tianxiao and Shi, Shuyuan and others},
  journal={Nano Letters},
  volume={24},
  number={23},
  pages={6931--6938},
  year={2024},
  publisher={ACS Publications}
}

@article{gupta2025,
  title={Harnessing orbital Hall effect in spin-orbit torque MRAM},
  author={Gupta, Rahul and Bouard, Chlo{\'e} and Kammerbauer, Fabian and Ledesma-Martin, J Omar and Bose, Arnab and Kononenko, Iryna and Martin, Sylvain and Us{\'e}, Perrine and Jakob, Gerhard and Drouard, Marc and others},
  journal={Nature Communications},
  volume={16},
  number={1},
  pages={130},
  year={2025},
  publisher={Nature Publishing Group UK London}
}

@article{Lyalin2023,
  title = {Magneto-Optical Detection of the Orbital Hall Effect in Chromium},
  author = {Lyalin, Igor and Alikhah, Sanaz and Berritta, Marco and Oppeneer, Peter M. and Kawakami, Roland K.},
  journal = {Phys. Rev. Lett.},
  volume = {131},
  issue = {15},
  pages = {156702},
  numpages = {6},
  year = {2023},
  month = {Oct},
  publisher = {American Physical Society},
  doi = {10.1103/PhysRevLett.131.156702},
  url = {https://link.aps.org/doi/10.1103/PhysRevLett.131.156702}
}

@article{Go2018,
  title = {Intrinsic Spin and Orbital Hall Effects from Orbital Texture},
  author = {Go, Dongwook and Jo, Daegeun and Kim, Changyoung and Lee, Hyun-Woo},
  journal = {Phys. Rev. Lett.},
  volume = {121},
  issue = {8},
  pages = {086602},
  numpages = {6},
  year = {2018},
  month = {Aug},
  publisher = {American Physical Society},
  doi = {10.1103/PhysRevLett.121.086602},
  url = {https://link.aps.org/doi/10.1103/PhysRevLett.121.086602}
}

@article{Bernevig2005,
  title = {Orbitronics: The Intrinsic Orbital Current in $p$-Doped Silicon},
  author = {Bernevig, B. Andrei and Hughes, Taylor L. and Zhang, Shou-Cheng},
  journal = {Phys. Rev. Lett.},
  volume = {95},
  issue = {6},
  pages = {066601},
  numpages = {4},
  year = {2005},
  month = {Aug},
  publisher = {American Physical Society},
  doi = {10.1103/PhysRevLett.95.066601},
  url = {https://link.aps.org/doi/10.1103/PhysRevLett.95.066601}
}

@article{ManchonROMP2019,
  title = {Current-induced spin-orbit torques in ferromagnetic and antiferromagnetic systems},
  author = {Manchon, A. and \ifmmode \check{Z}\else \v{Z}\fi{}elezn\'y, J. and Miron, I. M. and Jungwirth, T. and Sinova, J. and Thiaville, A. and Garello, K. and Gambardella, P.},
  journal = {Rev. Mod. Phys.},
  volume = {91},
  issue = {3},
  pages = {035004},
  numpages = {80},
  year = {2019},
  month = {Sep},
  publisher = {American Physical Society},
  doi = {10.1103/RevModPhys.91.035004},
  url = {https://link.aps.org/doi/10.1103/RevModPhys.91.035004}
}

@article{jo2024spintronics,
  title={Spintronics meets orbitronics: Emergence of orbital angular momentum in solids},
  author={Jo, Daegeun and Go, Dongwook and Choi, Gyung-Min and Lee, Hyun-Woo},
  journal={npj Spintronics},
  volume={2},
  number={1},
  pages={19},
  year={2024},
  publisher={Nature Publishing Group UK London}
}

@article{go2021orbitronics,
  title={Orbitronics: Orbital currents in solids},
  author={Go, Dongwook and Jo, Daegeun and Lee, Hyun-Woo and Kl{\"a}ui, Mathias and Mokrousov, Yuriy},
  journal={Europhysics Letters},
  volume={135},
  number={3},
  pages={37001},
  year={2021},
  publisher={EDP Sciences, IOP Publishing and Societ{\`a} Italiana di Fisica}
}

@article{cysne2025orbitronics,
  title={Orbitronics in two-dimensional materials},
  author={Cysne, Tarik P and Canonico, Luis M and Costa, Marcio and Muniz, RB and Rappoport, Tatiana G},
  journal={npj Spintronics},
  volume={3},
  number={1},
  pages={39},
  year={2025},
  publisher={Nature Publishing Group UK London}
}

@article{tanaka2008intrinsic,
	title={Intrinsic spin Hall effect and orbital Hall effect in 4 d and 5 d transition metals},
	author={Tanaka, T and Kontani, Hiroshi and Naito, Masayuki and Naito, T and Hirashima, Dai S and Yamada, K and Inoue, J-I},
	journal={Physical Review B—Condensed Matter and Materials Physics},
	volume={77},
	number={16},
	pages={165117},
	year={2008},
	publisher={APS}
}

@article{kontani2009giant,
	title={Giant orbital Hall effect in transition metals: Origin of large spin and anomalous Hall effects},
	author={Kontani, Hiroshi and Tanaka, T and Hirashima, DS and Yamada, K and Inoue, J},
	journal={Physical review letters},
	volume={102},
	number={1},
	pages={016601},
	year={2009},
	publisher={APS}
}

@article{jo2018gigantic,
	title={Gigantic intrinsic orbital Hall effects in weakly spin-orbit coupled metals},
	author={Jo, Daegeun and Go, Dongwook and Lee, Hyun-Woo},
	journal={Physical Review B},
	volume={98},
	number={21},
	pages={214405},
	year={2018},
	publisher={APS}
}

@article{choi2023observation,
  title={Observation of the orbital Hall effect in a light metal Ti},
  author={Choi, Young-Gwan and Jo, Daegeun and Ko, Kyung-Hun and Go, Dongwook and Kim, Kyung-Han and Park, Hee Gyum and Kim, Changyoung and Min, Byoung-Chul and Choi, Gyung-Min and Lee, Hyun-Woo},
  journal={Nature},
  volume={619},
  number={7968},
  pages={52--56},
  year={2023},
  publisher={Nature Publishing Group UK London}
}

@article{gao2025nonlocal,
	title={Nonlocal electrical detection of reciprocal orbital Edelstein effect},
	author={Gao, Weiguang and Liao, Liyang and Isshiki, Hironari and Budai, Nico and Kim, Junyeon and Lee, Hyun-Woo and Lee, Kyung-Jin and Go, Dongwook and Mokrousov, Yuriy and Miwa, Shinji and others},
	journal={Nature Communications},
	volume={16},
	number={1},
	pages={6380},
	year={2025},
	publisher={Nature Publishing Group UK London}
}

@article{rang2024orbital,
	title={Orbital relaxation length from first-principles scattering calculations},
	author={Rang, Max and Kelly, Paul J},
	journal={Physical Review B},
	volume={109},
	number={21},
	pages={214427},
	year={2024},
	publisher={APS}
}

@article{hayashi2023observation,
  title={Observation of long-range orbital transport and giant orbital torque},
  author={Hayashi, Hiroki and Jo, Daegeun and Go, Dongwook and Gao, Tenghua and Haku, Satoshi and Mokrousov, Yuriy and Lee, Hyun-Woo and Ando, Kazuya},
  journal={Communications Physics},
  volume={6},
  number={1},
  pages={32},
  year={2023},
  publisher={Nature Publishing Group UK London}
}

@article{xu2021ab,
	title={Ab initio ultrafast spin dynamics in solids},
	author={Xu, Junqing and Habib, Adela and Sundararaman, Ravishankar and Ping, Yuan},
	journal={Physical Review B},
	volume={104},
	number={18},
	pages={184418},
	year={2021},
	publisher={APS}
}

@article{xu2020spin,
	title={Spin-phonon relaxation from a universal ab initio density-matrix approach},
	author={Xu, Junqing and Habib, Adela and Kumar, Sushant and Wu, Feng and Sundararaman, Ravishankar and Ping, Yuan},
	journal={Nature communications},
	volume={11},
	number={1},
	pages={2780},
	year={2020},
	publisher={Nature Publishing Group UK London}
}

@article{xu2021giant,
	title={Giant spin lifetime anisotropy and spin-valley locking in silicene and germanene from first-principles density-matrix dynamics},
	author={Xu, Junqing and Takenaka, Hiroyuki and Habib, Adela and Sundararaman, Ravishankar and Ping, Yuan},
	journal={Nano letters},
	volume={21},
	number={22},
	pages={9594--9600},
	year={2021},
	publisher={ACS Publications}
}

@article{xu2023ab,
	title={Ab initio predictions of spin relaxation, dephasing, and diffusion in solids},
	author={Xu, Junqing and Ping, Yuan},
	journal={Journal of Chemical Theory and Computation},
	volume={20},
	number={2},
	pages={492--512},
	year={2023},
	publisher={ACS Publications}
}

@article{xu2024spin,
	title={How spin relaxes and dephases in bulk halide perovskites},
	author={Xu, Junqing and Li, Kejun and Huynh, Uyen N and Fadel, Mayada and Huang, Jinsong and Sundararaman, Ravishankar and Vardeny, Valy and Ping, Yuan},
	journal={Nature Communications},
	volume={15},
	number={1},
	pages={188},
	year={2024},
	publisher={Nature Publishing Group UK London}
}

@article{multunas2023circular,
	title={Circular dichroism of crystals from first principles},
	author={Multunas, Christian and Grieder, Andrew and Xu, Junqing and Ping, Yuan and Sundararaman, Ravishankar},
	journal={Physical Review Materials},
	volume={7},
	number={12},
	pages={123801},
	year={2023},
	publisher={APS}
}

@article{grieder2025relation,
	title={Relation of Continuous Chirality Measure to Spin and Orbital Polarization, and Chiroptical Properties in Solids},
	author={Grieder, Andrew and Tu, Shihao and Ping, Yuan},
	journal={Advanced Optical Materials},
	volume={13},
	number={33},
	pages={e01190},
	year={2025},
	publisher={Wiley Online Library}
}

@article{thonhauser2005orbital,
  title={Orbital magnetization in periodic insulators},
  author={Thonhauser, Timo and Ceresoli, Davide and Vanderbilt, David and Resta, Raffaele},
  journal={Physical review letters},
  volume={95},
  number={13},
  pages={137205},
  year={2005},
  publisher={APS}
}

@article{cysne2022orbital,
  title={Orbital Hall effect in bilayer transition metal dichalcogenides: From the intra-atomic approximation to the Bloch states orbital magnetic moment approach},
  author={Cysne, Tarik P and Bhowal, Sayantika and Vignale, Giovanni and Rappoport, Tatiana G},
  journal={Physical Review B},
  volume={105},
  number={19},
  pages={195421},
  year={2022},
  publisher={APS}
}

@article{simoni2025,
    author = {Simoni, Jacopo and Riva, Gabriele and Ping, Yuan},
    title = {First-principles open quantum dynamics for solids based on density-matrix formalism},
    journal = {The Journal of Chemical Physics},
    volume = {163},
    number = {17},
    pages = {170901},
    year = {2025},
    month = {11},
    issn = {0021-9606},
    doi = {10.1063/5.0277603},
    url = {https://doi.org/10.1063/5.0277603},
}

@article{pezo2023orbital, title={Orbital Hall physics in two-dimensional Dirac materials}, author={Pezo, Armando and Garc{'\i}a Ovalle, Diego and Manchon, Aur{'e}lien}, journal={Physical Review B}, volume={108}, number={7}, pages={075427}, year={2023}, publisher={APS} }

@article{gridnev2001,
	author = {V.N. Gridnev},
	title = {{Theory of Faraday Rotation Beats in Quantum Wells with Large Spin Splitting}},
	journal = {JETP Letters},
	volume = {74},
	pages = {380--383},
	year = {2001},
	doi = {https://doi.org/10.1134/1.1427126},
}

@article{Brand2002,
	author =
	{
		Brand, M. A.
		and Malinowski, A.
		and Karimov, O. Z.
		and Marsden, P. A.
		and Harley, R. T.
		and Shields, A. J.
		and Sanvitto, D.
		and Ritchie, D. A.
		and Simmons, M. Y.
	},
	title = {{Precession and Motional Slowing of Spin Evolution in a High Mobility Two-Dimensional Electron Gas}},
	journal = {Phys. Rev. Lett.},
	volume = {89},
	pages = {236601},
	year = {2002},
	doi = {10.1103/PhysRevLett.89.236601},
}

\end{document}